# Reversible control of spin-polarised supercurrents in ferromagnetic Josephson junctions

N. Banerjee[*], J. W. A. Robinson and M. G. Blamire

Department of Materials Science and Metallurgy, University of Cambridge, 27 Charles Babbage Road, Cambridge CB3 0FS, United Kingdom.

[*] nb366@cam.ac.uk

Magnetic inhomogeneity at a superconductor (S) – ferromagnet (F) interface converts spin-singlet Cooper pairs into spin-one triplet pairs. These pairs are immune to the pair-breaking exchange field in F and support a long-range proximity effect. Although recent experiments have confirmed the existence of spin-polarised triplet supercurrents in S-F-S Josephson junctions, reversible control of the supercurrent has been impossible because of the robust pre-configured nature of the inhomogeneity. Here we use a barrier comprising three F layers whose relative magnetic orientation, and hence the interfacial inhomogeneity, can be controlled by small magnetic fields; we show that this enables full control of the triplet supercurrent and, by using finite element micromagnetic simulations, we can directly relate the experimental data to the theoretical models which provide a general framework to understand the role played by magnetic states in long-range supercurrent modulation.

## Introduction

The interplay between superconducting and magnetic order parameters constrained by the exclusion principle and fermionic exchange statistics has given rise to rich and diverse physics and reignited the interest in the problem of coexistence of magnetism and superconductivity[1,2]. Of particular relevance is the theoretical prediction[3] that magnetic inhomogeneity at a S-F interface leads to a conversion between singlet and triplet spin pairing states in different quantization bases and produces equal spin Cooper pairs. Recent experimental verification[4–14] of long-ranged supercurrents in ferromagnets has raised the intriguing possibility of taking the next step towards practical implementation as a dissipation-less version of spin electronics (spintronics)[15]. Two key aspects need to be addressed for a realisation of such circuits: efficient generation of spin-polarised supercurrents, and their active control. Efforts[16–18] in the last few years have been primarily directed towards optimising the supercurrent; little progress has so far been made in directly controlling it. Optimising the inhomogeneity in the form of a robust spin-mixer layer which maximises the singlet to triplet conversion ironically appears to make it difficult to design an externally controllable system.

In this article, we report SF'FF'S Josephson junctions in which the magnetic alignment between thin F' mixer layers (composed of the soft ferromagnet $Ni_{80}Fe_{20}$, Permalloy, Py) and a thicker F layer (Co) can be controlled by the applied magnetic field ($H$) and show that the magnitude of the critical current $I_C$ is controlled by the net misalignment of the magnetism in the three layers. In particular, we show that the supercurrent is zero for the parallel aligned case (Fig. 1a,b). This device is the superconducting analogue of the spin valve which is the foundation of conventional spintronics[19]. We analyse our results based on the Houzet and Buzdin model[20] of a Josephson junction incorporating a trilayer magnetic structure which, in combination with finite element analysis, provides a semi-quantitative fit to the data.

## Results

### *Transport measurements of SF'FF'S Josephson junctions*

Figure 2(a) shows the $I_C$ in a Josephson junction with a Py(1.5)/Cu(5)/Co(5.5)/Cu(5)/Py(1.5) (thicknesses in nanometres) barrier. The behaviour is distinctly different from the expected dependence of $I_C$ on $H$ ($I_C(H)$) in a SFS Josephson junction: an example is shown in Fig. 2(b) from a junction incorporating Ho mixer layers at the S/F interface but having the same Co layer thickness (6 nm) and comparable dimensions in which, although hysteretic, $I_C(H)$ clearly follows the expected Fraunhofer-type dependence with distinct second lobes. In Fig. 2(a) and, from a similar device, in Fig.

3(a) we observe that the overall shape and behaviour of $I_C(\mathbf{H})$ is very different: $I_C$ goes to zero above a certain field magnitude instead of showing multiple oscillations with field (confirmed by the linear current-voltage characteristic recorded at -40 mT (Fig. 3(a) inset)); the small rise seen at high fields is associated with thermal effects arising from the magnet coil. On reducing the central Co layer thickness to 3 nm which enables a singlet contribution to the supercurrent, although the central peak remains strongly distorted, additional lobes reappear beyond the first minima (Fig. 2(c)).

Although unusual $I_C(\mathbf{H})$ patterns have been reported before in SFS, SF'FF'S or SIFS Josephson junctions[16,21–23]; these are irreproducible and attributed to stochastic variations of the flux arising from a multi-domain magnetic barrier. This is distinctly different from what we observe here: a highly reproducible but strongly distorted central peak with zero critical current beyond a certain magnetic field value.

We start the discussion of these results by setting an upper limit for the singlet $I_C$ through such devices. In general the singlet $I_C$ in SFS Josephson junctions will be oscillatory with multiple $0 - \pi$ transitions with increasing F thickness[24,25], but to provide an estimate of the upper limit of the singlet current we just consider the envelope of the $I_C$ maxima – in other words assuming that the net exchange energy of the barrier is such that singlet pair dephasing is zero and that the supercurrent is just limited by the coherence lengths. The singlet coherence lengths $\xi_{Co}$ and $\xi_{Py}$ have been measured to be 3.0 nm and 1.4 nm respectively[26], meaning that the total F barrier thickness is equivalent to 12 nm of Co for the devices shown in Fig. 2(a) and 3(a). Taking a typical value of of 1.5 µV as the characteristic voltage ($I_C R_N$) in such junctions (extrapolated for 12 nm Co thickness) previously observed for Nb/Co/Nb[25], gives a maximum singlet $I_C$ of ~ 40 µA. This value does not take into account the additional scattering at the multiple interfaces in our structures[27]. To take account of these interfaces and at least partial cancellation of the dephasing, a more representative number might be obtained by extrapolating from similar sized junctions with much thinner Py(1.6)/Cu(8)/Co(1) ferromagnetic barriers[28], for which the $I_C R_N$ varied between 0.8-2 µV thus, giving a maximum $I_C$ of ~ 60 µA when the two F layers were AP. In our devices the $I_C R_N$ ranged from 4-11 µV with a corresponding $I_C$ of ~ 500-600 µA. We therefore conclude that the supercurrents cannot originate from singlet pair transport and so must be primarily mediated by spin-one triplet pairs.

The non-collinearity between the adjacent F layers which is required for triplet generation[20] arises from the complex magnetic microstructure of the F layers which itself is due to a competition between the dipolar field[29–31], magnetic anisotropies and the external field. Since this microstructure changes with the applied field, the maximum critical current, $I_{C0}$ should depend on $\mathbf{H}$.

For our junctions, the $I_C(\mathbf{H})$ modulation is controlled by two factors: firstly, the field-dependent magnetic inhomogeneity determines the maximum triplet supercurrent $I_{C0}$ and, secondly, phase-variations arising from the applied field and induced changes to the net barrier moment determine the net $I_C$ (which leads to the Fraunhofer $I_C(\mathbf{H})$ modulation seen in conventional junctions). Both of these factors depend on the details of the micromagnetic configuration of each magnetic layer and so it is necessary to understand how this depends on $\mathbf{H}$.

### *Finite element analysis of $I_C$ dependence on magnetic field*

Experimentally it is hard to directly visualise these states in sub-micron devices and, although previous SFS experiments have used indirect information from magnetic measurements of unpatterned films, the much weaker role of dipolar fields in continuous films means that it is impossible to directly relate the details of the micromagnetic structure of nanopillar devices from such measurements. Instead we have used finite element micromagnetic simulations[32] using Object Oriented Micromagnetic Framework (OOMMF) which allows simulation of the magnetic state up to a resolution of few nanometres and make semi-quantitative predictions relating the magnetic structure to the spin-polarised supercurrent flowing through the device. The saturation magnetization, exchange coefficient and uniaxial anisotropy for Co were set to $1400 \times 10^3$ Am$^{-1}$, $3 \times 10^{-11}$ Jm$^{-1}$ and $208 \times 10^3$ Jm$^{-3}$ respectively while for Py these values were, $860 \times 10^3$ Am$^{-1}$, $1.3 \times 10^{-11}$ Jm$^{-1}$ and 150 Jm$^{-3}$ respectively. The saturation magnetization and exchange coefficient values were taken from the OOMMF database (used in the literature); the saturation magnetization agrees closely with the values we have calculated from bulk films deposited on SiO$_2$ substrates sandwiched between 100 nm thick Cu. The uniaxial anisotropy for Py was calculated from the difference in area of the hysteresis loops measured along the hard and easy axis and the direction, originally set by the growth field, was orthogonal to the applied field $H$. However, it is seen that the dipolar energy term in this case is much larger than the Py anisotropy energy and alone dictates the ground state configuration. To determine the value of Co anisotropy we have simulated a spin valve structure consisting of Co (1.5)/Cu (7.5)/Py (1.5) similar to the one used in Ref[28] which was grown under similar conditions and modified the Co anisotropy value to match the switching field obtained from magnetoresistance measurements on these structures. The value obtained from these simulations is 40% of the reported value in OOMMF database; this is not unexpected since the anisotropy strongly depends on the growth conditions, the substrate used and the film thickness[33]. The Co anisotropy was in the plane of the layer and the direction was chosen from a random vector field which reflects the polycrystalline nature of the sputtered films. The damping coefficient was set to 0.5 which allowed for rapid convergence.

Figure 3(a) shows one branch (positive to negative field sweep) of $I_C(\boldsymbol{H})$ for a device. Micromagnetic simulations for this device have been performed at 5 mT intervals for an equivalent field sweep: Fig. 3(c) shows plan views of the magnetic structure of each layer at representative fields. The colour scheme adopted to represent magnetization direction is red-white-blue with red (blue) pixels representing magnetization aligned along the positive (negative) external field direction. White pixels represent magnetic moments orthogonal to the applied field direction. At the highest field magnitudes, the three F layers are parallel. Around 10 mT, the Py layers start inhomogeneously reversing under the dipolar magnetostatic interaction from the Co layer and are fully reversed at zero field. As the field increases in the negative direction the Co layer eventually reverses beyond -10 mT. It is clear from the images that significant non-collinearity exists *within* and *between* all layers during the reversal process.

This observation is important in its own right as there have been speculations about the specific origin of spin-polarised supercurrents in SF'FF'S devices before. Although it was concluded by Khasawneh *et al.*[22] that non-collinearity between F' and F layers most likely gives rise to the spin-polarised supercurrents rather than inhomogeneity in F' layers, our simulations indicate a more subtle effect at play. Intuitively one might be inclined to believe that there is little inhomogeneity in nano-pillar devices, but it is evident here that inhomogeneity does exist and if engineered properly using F layer with difference in coercivities, this can be translated to a *local* non-collinearity between Py and Co layers which is critical for spin-polarised supercurrent generation.

To proceed further, a quantitative estimate of the magnetic inhomogeneity as a function of $\boldsymbol{H}$ is required to estimate of the $I_{C0}$ through the junction. According to the Houzet-Buzdin model, $I_{C0}$ for a Josephson junction at a fixed temperature with a $F_1'FF_2'$ barrier is proportional to the product of the sines of the angles between adjacent magnetic layers ($\phi_1$ and $\phi_2$), i.e.

$$I_{C0} \propto \sin\phi_1 \sin\phi_2. \quad (1)$$

Since the F layers cannot be approximated by a macrospin, it implies we have to apply the model by calculating the product of the sine of the angle between the cells of two adjacent magnetic layers for each vertical cell stack used in the simulations within which a continuum approximation implies a uniform magnetization. The components of the magnetization in each cell are known from the OOMMF simulation and $\sin\phi_1$ is obtained from the inner product of the magnetization in the $i^{th}$ cell of the top Py with the corresponding cell in Co. The same procedure is repeated for the $i^{th}$ cell of the bottom Py and Co to obtain $\sin\phi_2$. The product $\sin\phi_1 * \sin\phi_2$ indicates the combined inhomogeneity arising from the three F layers (outer Py layers and the central Co layer). This procedure is repeated for each cell in the entire layer and an average value of $\sin\phi_1 * \sin\phi_2$ is obtained by summing the product for all the cells and dividing by the total number of cells. We have taken into account the actual sign of the product $\sin\phi_1 * \sin\phi_2$, since according to the Houzet-Buzdin model the junction can be in a 0 ($\pi$) state depending on the anti-parallel (parallel) orientation of the magnetization of the outer layers. This is clear from the micromagnetic simulations shown in Fig. 3c; at low fields, local regions of the junctions are in a 0 or $\pi$ state thereby reducing the total critical current through the junction. Also, the dependence of the critical current on the relative angle between two F layers reflects the fact that non-collinearity induced by inhomogeneity between two F layers is more important than inhomogeneity in a single F layer where it occurs at the scale of the magnetic exchange length, which far exceeds the coherence length of a Cooper pair in the F layer.

Figure 3(b) (inset) shows the dependence of $I_{C0}$ on **H**: there are two distinct peaks (indicating maximum inhomogeneity), with the first peak at a positive field (~10mT) related primarily to Py reversal while the second (~-10mT) is due to the Co layer reversal.

To calculate the phase-variation owing to the local flux density **B** arising from a combination of the inhomogeneous barrier magnetization and **H**, we integrate the variation of the phase difference of the superconducting order parameter ($\phi$) over the junction area

$$I_C = I_{C0} \int_S \sin\left[\phi_0 + \int_a \left\{\frac{2e}{\hbar}\left(\int_{\lambda-\frac{d}{2}}^{\lambda+\frac{d}{2}} \boldsymbol{B} dz\right) \times \hat{\boldsymbol{z}}\right\} \cdot d\boldsymbol{l}\right] dS, \qquad (2)$$

where $\lambda$ is the penetration depth of the superconductor, $\hat{\boldsymbol{z}}$ is the direction normal to the plane of the junction and $e$ is the charge of an electron. Here, the line integral is carried out for all the points $a$ defining the junction by starting from the origin where $\phi_0$ is defined. The critical current is finally obtained by maximising with respect to $\phi_0$ of the surface integral defining the junction over the points $a$. The effective value of $\lambda$ for our materials and geometry is estimated to be 90 nm (by measuring the field corresponding to one flux quantum) from devices with similar dimensions and Nb thicknesses but with Ho as the triplet generators instead of Py (Fig. 2(b)). Given the complex magnetization distribution in our junctions, a simple analytical solution to equation (2) is not possible and so we apply a numerical technique[34] as outlined below:

The local $\boldsymbol{B}$ fields obtained from the micromagnetic simulations when integrated vertically normal to the plane of the layers for the whole barrier thickness including the London penetration depth ($\lambda$) of the Nb electrodes, gives a linear flux density matrix ($\boldsymbol{b}_{ij}$) according to

$$\boldsymbol{b}_{ij} = \int_{\lambda-\frac{d}{2}}^{\lambda+\frac{d}{2}} \boldsymbol{B} dz. \quad (3)$$

Here $z$ is the direction normal to the plane of the films. The $\boldsymbol{b}_{ij}$ matrix is then converted to an equivalent matrix of phase gradients ($\boldsymbol{\phi}'_{ij}$) according to

$$\boldsymbol{\phi}'_{ij} = \boldsymbol{b}_{ij} \times \hat{\boldsymbol{z}}. \quad (4)$$

The critical current is then obtained by performing the summations

$$I_C = I_{C0} \sum_x \sum_y \sin\left(\phi_0 + \frac{2e}{\hbar}\sum_{i=1}^{x} \boldsymbol{\phi}'_{i1} \cdot \delta\boldsymbol{x} + \frac{2e}{\hbar}\sum_{j=1}^{y} \boldsymbol{\phi}'_{xj} \cdot \delta\boldsymbol{y}\right). \quad (5)$$

and maximizing with respect to $\phi_0$, where $\phi_0$ is set at (1,1).

In order to compare our simulations with experimental data it is necessary to know the effective coupling of the flux originating from the barrier magnetization into the junction. In an SFS junction with a single, homogeneously magnetised ferromagnetic barrier the maximum critical current is achieved when $\boldsymbol{H} = \Delta\boldsymbol{H}$ where

$$\Delta\boldsymbol{H} (2*\lambda + d_{\text{Co}}) = b\mu_0 \boldsymbol{M} d_{\text{FM}}, \quad (6)$$

$b$ is the effective coupling of the flux originating from the saturation magnetization of the ferromagnet $\boldsymbol{M}$, and $d_{\text{FM}}$ is the thickness of the ferromagnet. We can estimate $b$ from Nb/Ho/Co/Ho/Nb junctions with a similar size and shape: the inset to Fig. 3b in Ref. 4 shows the field offset ($\Delta\boldsymbol{H}$) vs $d_{\text{Co}}$ and a linear fit to ($\Delta\boldsymbol{H}$ vs $d_{\text{Co}}/(2*\lambda + d_{\text{Co}})$ (Fig. 3B inset, Ref 4) gives $b = 0.2$. This implies a significant partial cancellation of the magnetisation flux arising from fringing fields producing a flux in the opposite direction in the region within the penetration depth of the superconductor. Using $b = 0.2$, $I_C$ calculated from equation (5) with $I_{C0} = 1$ is shown in Fig. 3(b) (inset); the distortion of an ideal Fraunhofer pattern arises due to field-dependent inhomogeneous magnetism of the barrier. Figure 3(b) (green curve) shows the full solution of equation (5) by including $I_{C0}(\boldsymbol{H})$ shown in the other inset and thus shows the combined effect of the dependence of the triplet supercurrent on the magnetic structure and the flux arising from the inhomogeneous magnetic barrier.

**Discussion**

Inspection of Fig. 3(b) shows that several features of the experimental curve are well reproduced. These include the rapid decay of $I_C$ above a critical field and the severely suppressed higher order lobes as a result of a more homogenous magnetic structure where $I_{C0} \to 0$. The small remnant oscillations in the simulated curve arises from a residual inhomogeneity at the edges arising from dipolar fields between Py and Co layers which always remain in the simulation but, experimentally may not contribute because of surface oxidation and intermixing arising from the ion-milling during fabrication. The dip near zero field (less prominent in the experimental curve) is quite sensitive to the magnetic configuration of each layer. At low fields, such configurations are quite prone to stochastic variations induced by factors like the film microstructure, exact device dimensions and magnetic history, and direct comparison with experiments are difficult to make in that field region. Ideally, the low and zero field configuration is expected to be symmetric with respect to the magnetic state of the outer layers and noncolinear to the central Co layer; this implies that globally the junction is in a $\pi$ state and the sign of the product of $\sin\phi_1 \sin\phi_2$ is irrelevant. Taking this fact into consideration, we have simulated the same junction (Fig. 3b, brown curve). The two simulations differ only at or near zero field. This brown curve, therefore, provides an upper limit to the critical current close to zero field for a junction with homogeneously symmetric (or antisymmetric) outer Py layers.

In view of the above, the behaviour of devices with thinner Co (Fig. 2(c)) becomes clear: on reducing the central Co layer a background singlet current flows whose maximum value is insensitive to the magnetic state in the device and is thus visible as phase-controlled $I_C$ oscillations beyond the central lobe.

From the point of view of applications, the key aspect of this result is the experimental proof that the triplet supercurrent amplitude can be reversibly controlled by changing the magnetic inhomogeneity within the barrier. This provides direct control over the spin-polarised supercurrent which is of fundamental importance towards the realisation of practical superconducting spintronic circuits. Perhaps equally as importantly, we demonstrate that significant inhomogeneity can be generated even in nanoscale junctions and appropriate engineering of the micromagnetic structure offers the potential to optimise the response of the system to very small field changes or spin transfer torques[35–37].

**Methods**

**Film growth.** Nb(250)/Cu(5)/Py(y)/Cu(5)/Co(x)/Cu(5)/Py(y)/Cu(5)/Nb(250) (thicknesses in nanometers) samples were grown on unheated (001) Si substrates with a 250 nm thick $SiO_2$ coating by dc magnetron sputtering in ultra-high vacuum chamber. The base pressure was maintained below $10^{-8}$ Pa while the chamber was cooled via a liquid nitrogen jacket. The targets were pre-sputtered for 15-20 minutes to clean the surfaces and the films were grown in 50 mT (approx.) magnetic field by placing the substrates between two bar magnets. This induces an easy axis for the Py films along the growth-field direction. The Cu layer between the base Nb and Py was inserted to eliminate exchange coupling. The Co thickness (x) was varied between 3 and 9 nm.

**Device fabrication.** Devices were prepared with either 1.5 or 2.5 nm Py layers (y); in general these showed similar results. Standard optical lithography and Ar-ion milling were used to define 4-µm wide tracks which were narrowed down by focused-ion-beam milling to make current-perpendicular-to-plane devices: details of the process are described elsewhere[38]. The average device dimensions were in the range of 600 nm × 500 nm.

**Transport measurements.** A custom-built liquid He dip probe was used to cool the devices down to 4.2K by dipping it in a liquid He dewar. Current-voltage characteristics were measured by a 4-point technique using a current-biased circuit attached to a lock-in amplifier. The Josephson effect in the devices was measured by applying an in-plane magnetic field and measuring the critical current $I_C$ as a function of the applied field ($H$) (Fig. 2a). The critical current was determined using a voltage criterion and hence a finite value is recorded even in the absence of a supercurrent. To subtract this background contribution, we have divided this criterion voltage by the normal state resistance of the junction which shifts the effective zero critical current line to the values shown by the red dotted line in each figure. The field was applied perpendicular to the Py easy axis which gives a weak tendency of the Py to align itself perpendicular to the Co layer at low or zero external fields.

**Acknowledgements**

M.G.B. acknowledges funding from the UK EPSRC and the European Commission through an ERC Advanced Investigator Grant "Superspin". J.W.A.R. acknowledges funding from the Royal Society.

**Author Contributions**

N.B. prepared the samples, fabricated the devices and carried out the measurements. N.B. performed the micromagnetic simulations and together with M.G.B. carried out the finite element simulations. N.B., M.G.B. and J.W.A.R. wrote the manuscript.

**Competing financial interests**

The authors declare no competing financial interests.

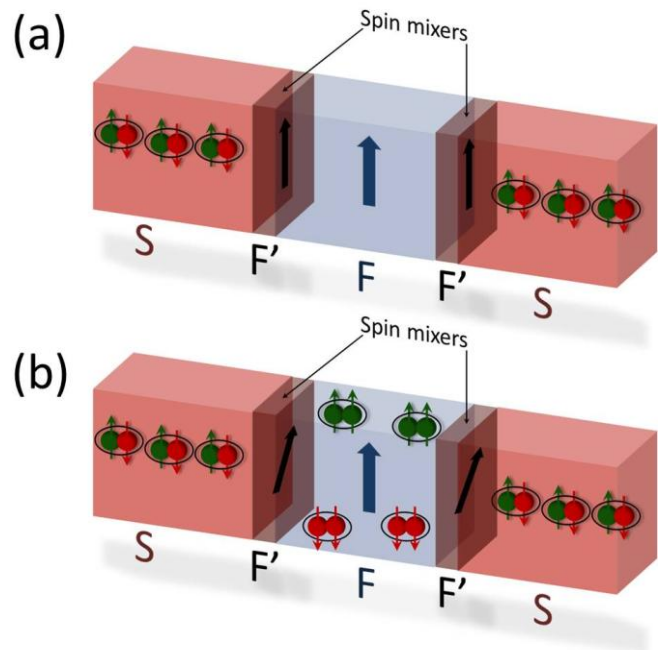

**Figure 1 | SF'FF'S Josephson junction containing a trilayer ferromagnet. (a)** At high magnetic fields the F layers are parallel (or anti-parallel) and the combined F layer thickness is much greater than the coherence length of the singlet Cooper pairs; no supercurrent flows through the structure. **(b)** At zero or low magnetic fields the inhomogeneous or non-collinear F' layers converts the spin-singlet Cooper pairs in S to equal spin-triplet Cooper pairs in F thus, allowing a finite triplet supercurrent to flow through the structure.

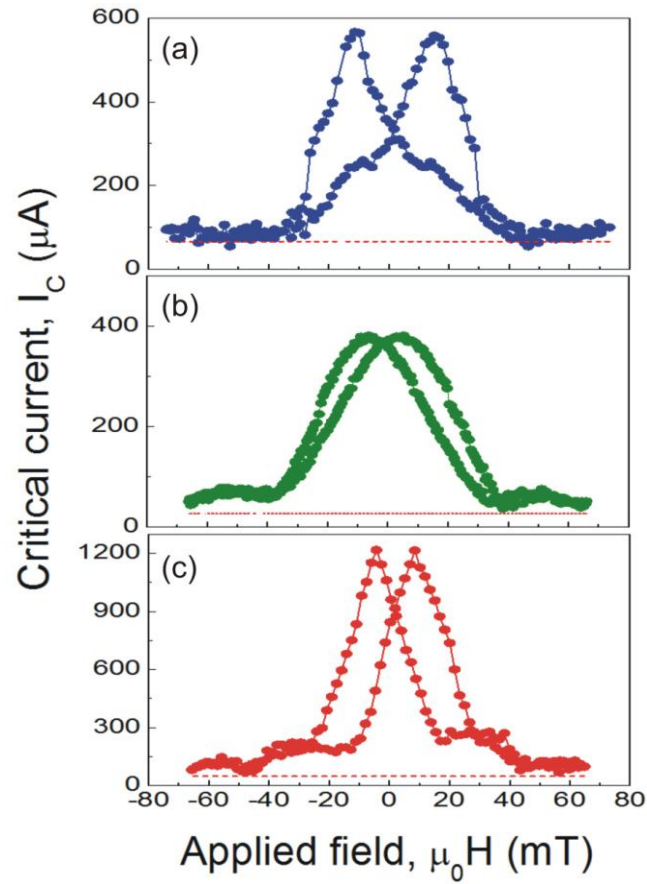

**Figure 2 | Dependence of the critical current on applied magnetic field of a Josephson junction**. **(a)** The junction is composed of a stack of Nb(250 nm)/Cu(5 nm)/Py(1.5 nm)/Cu(5 nm)/Co(5.5 nm)/Cu(5 nm)/Py(1.5 nm)/Cu(5 nm)/Nb(250 nm). **(b)** Josephson junction with 6 nm central Co layer but having 4.5 nm thick Ho layers at Nb/Co interface instead of Py to generate spin-polarised supercurrents. It shows a Fraunhofer-like dependence of the junction critical current with prominent side lobes. **(c)** A Josephson junction having a layer sequence similar to (a) but with a reduced central Co thickness of 3 nm showing oscillations of the critical current beyond the first lobe. The approximate dimensions of all the junctions are 600 nm X 500 nm. The red dotted lines in **a, b** and **c** represent the shift in the zero critical current line due to the finite non-zero voltage used to measure the critical current (see Methods section for details).

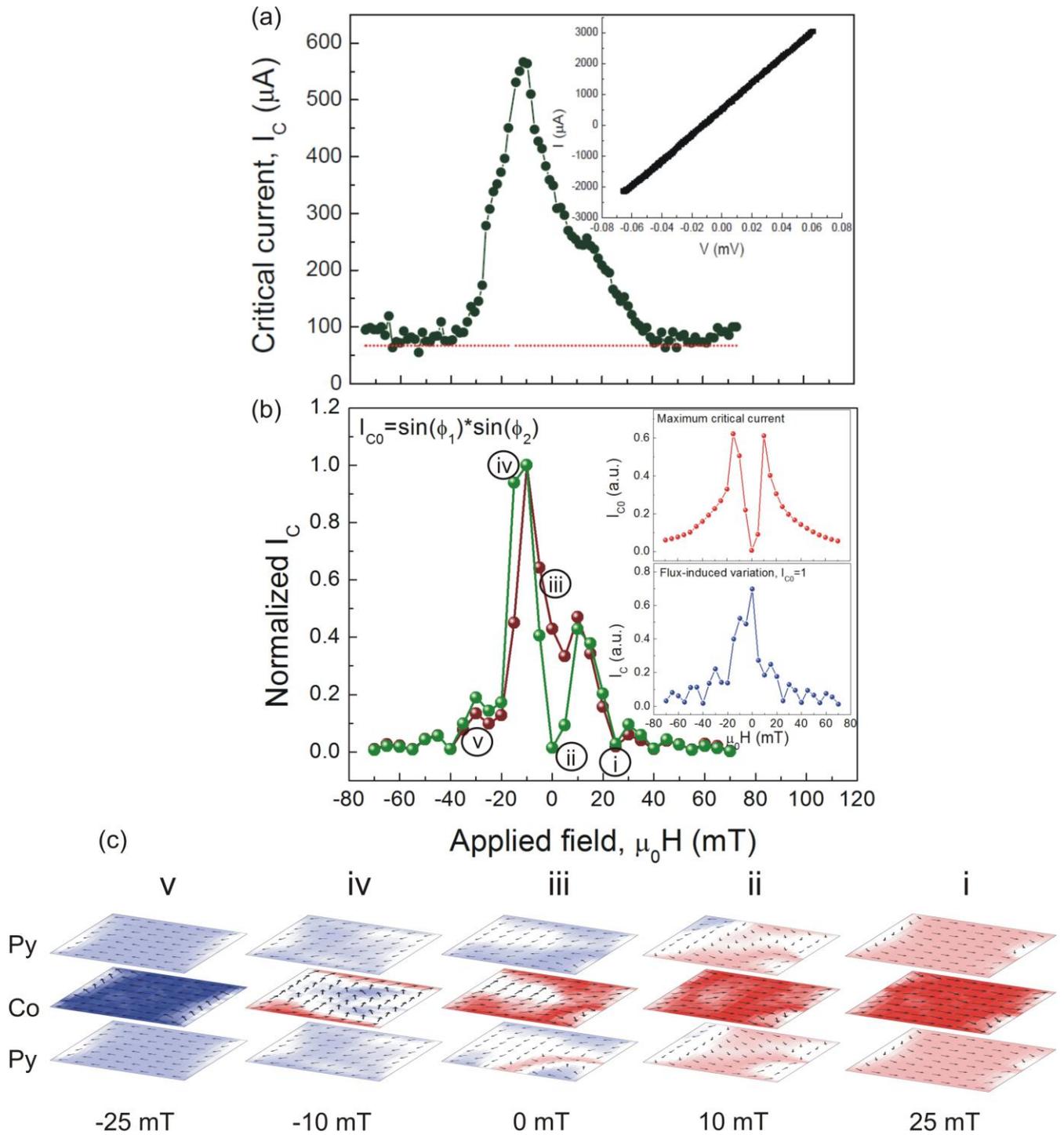

**Figure 3 | Experimental and simulated critical current variation with in-plane magnetic field (a)** Critical current versus in-plane magnetic field of a Nb/Cu(5 nm)/Py(1.5 nm)/Cu(5 nm)/Co(5.5 nm)/Cu(5 nm)/Py(1.5 nm)/Cu(5 nm)/Nb junction measured at 4.2 K. The red dotted line in **a** representing the shift in the zero critical current line due to the finite non-zero voltage used to measure the critical current (see Method section for details). The inset shows the current-voltage characteristic of the junction recorded at -40 mT to verify the absence of critical current. **(b)** Simulated $I_C(H)$ pattern (green and brown curves) showing the combined effect of inhomogeneous

magnetic state giving rise to a spin-polarised supercurrent and the effect of the flux taking into account the magnetic inhomogeneity. The green curve takes into account the actual sign of $\sin\phi_1 \sin\phi_2$ and thus accounts for the sign of the supercurrent depending on local $0$ or $\pi$ states whereas the brown curve only takes the modulus of $\sin\phi_1 \sin\phi_2$ . Inset shows the variation of the maximum supercurrent ($I_{C0}$) in the junction and the combined effect of flux arising from an inhomogeneous barrier moment and the applied field on the critical current as a function of an in-plane applied magnetic field. **(c, i-v)** The plan views of the magnetic states (from OOMMF simulations) for outer Py and central Co layers are shown with the corresponding magnetic fields as indicated below. The states corresponding to the field values shown are also marked in Fig. 3(b).